\documentclass[a4paper, 11pt]{article}
\usepackage{epsf}
\usepackage{epsfig}
\usepackage{psfrag}
\usepackage{fancyhdr}
\usepackage{amsmath}
\usepackage{amssymb}
\usepackage{theorem}
\numberwithin{equation}{section}

\begin{document}
\title{Relevance Vector Machines for classifying points and regions in biological
  sequences.}
\author{Thomas A. Down\footnote{Corresponding author} and Tim J. P. Hubbard\\
Wellcome Trust Sanger Institute, Hinxton, Cambridge, CB10 1SA\\
{\it \{td2,th\}@sanger.ac.uk}}
\date{\today}
\maketitle

\newpage

\begin{abstract}
The Relevance Vector Machine (RVM) is a recently developed machine learning framework capable
of building simple models from large sets of candidate features.  Here, we
describe a protocol for using the RVM to explore very large numbers of candidate
features, and a family of models which apply the power of the RVM to classifying
and detecting interesting points and regions in biological sequence data.  The
models described here have been used successfully for predicting transcription
start sites and other features in genome sequences.
\end{abstract}

\newpage

\section{Background}

Machine learning technologies have played an important role in bioinformatics for some years
now, both for extracting information from existing data sets and for making
predictions of interesting features in previously unannotated data
\cite{baldi.learning}.  Important
examples include Hidden Markov Models for annotating protein domains \cite{sonnhammer.pfam} and
neural networks for predicting protein secondary structure \cite{qian.secondary}.  Here, we describe a 
family of new machine learning methods for classifying biological sequences
based on the occurrence of particular motifs within the sequence.  All our
methods share some common features: they can all be represented as generalized
linear models and trained using the Sampling Relevance Vector Machine strategy,
and they all use position-weight matrices (PWMs) as the basic sensors to detect
motifs in the sequence data.  However, differences in the way that that PWM
scores are presented to the linear model provide us with several learning
methods suited to different purposes.

In this paper, we first review the basic Relevance Vector Machine machine method,
and describe some extensions which make the RVM easier to apply to complex
problems.  We then describe three different RVM-sequence models, one for
analyzing individual points in sequences ({\it e.g.} transcription start sites,
splice sites) and two for analyzing larger regions of sequence.  These methods
have been applied to a range of interesting sequence analysis questions, and we
highlight some applications throughout the paper.

\subsection {Generalized Linear Models}

Generalized linear models (GLMs) are a commonly used form of model for both classification
(separate data into two or more classes) and regression (estimate the value of
a continuous function) problems.  GLMs take the form:
\begin{equation}
\eta({\mathbf{x}}) = \sum^{M}_{m=1} \beta_m \phi_m ({\mathbf{x}}) + K
\end{equation}
where $\phi$ is a set of $M$ basis functions (which can be arbitrary real-valued functions)
and $\beta$ is a vector of weights.  While, in the machine learning literature,
the data items are usually vectors in some data space, since the model only refers
to the data via the basis functions, any type of data can be used so long as
appropriate basis function are provided.  The classical view of `training' a GLM is simply
to find some weights which offer a good fit to the provided training data.  Since
in this case we are considering classification  problems, we apply the 
logistic link function to the model output:
\begin{equation}
\pi({\mathbf{x}}) = \frac {1} {1 + e^{-\eta({\mathbf{x}})}}
\end{equation}
This renormalizes the model output such that $0 \leq \pi({\mathbf{x}}) \leq 1$, and can be interpreted
as a probability that ${\bf x}$ is a member of the `positive' class for the
classification problem in hand.  The classical problem of learning a set of weights,
$\beta$, such that the model optimally classifies some set of labeled training
data has been called logistic regression, and can be solved using techniques
such as Iterated Reweighted Least Squares \cite{nabney.irls}.

One way of looking at generalized linear models is that
the basis functions define a projection of the data into a high-dimensional
space (called feature space) where the data is either linear (for regression
problems) or linearly separable (for classification problems).  Clearly, the
choice of projection is important.  An important step in GLM learning
is to find a feature space which allows a linear model to fit the training
data, while not being of such high dimensionality that overfitting becomes
a problem.  Processes for solving this problem are described as feature selection.

\subsection {The Relevance Vector Machine}

The Relevance Vector Machine (RVM) \cite{tipping.rvm} is a sparse method for training
generalized linear models.  This means that it will generally select only a subset
(often a small subset) of the provided basis functions to use in the final model.
Sparsity is generally considered a desirable feature in a machine learning system
\cite{graepel.sparse}.  This is consistent with the intuition that a 
simple model is more likely to make useful generalizations which can be applied
to unseen data, rather than solving the problem `trivially' by memorizing the training data.
Sparse training methods also offer a partial solution to the question of feature selection.  The RVM
is named by analogy to the better-known Support Vector Machine method, which is also
a kind of sparse GLM trainer, and indeed the RVM was initially presented as an alternative and
direct competitor to the SVM.  However, SVMs can only be applied to training a restricted
subset of generalized linear models -- those that can be defined by a suitable
kernel function -- while the RVM can train a GLM with any collection of basis
functions.  The ability to perform sparse training with arbitrary basis functions
opens up many new possibilities, and is crucial for all the sequence analysis
methods described below.

Briefly, the Relevance Vector Machine is a Bayesian approach for training a
linear model.  In a binary classification problem, where each training datum ${\bf x}_n$
has a label $t_n$ (either $0$ or $1$), we can represent the probability that the
data set is correctly labeled given some classifier model $\pi({\bf x})$ as:
\begin{equation}
  \label{rvm.likelihood}
P(t | {\bf x}, \beta) = \prod_{n=1}^N \pi({\bf x}_n)^{t_n}(1-\pi({\bf x}_n))^{1-t_n}
\end{equation}
Where $\beta$ is the weights vector of our classification GLM.  If we assume that
the training data is correctly labeled, Bayes' theorem allows us to turn this
expression around and infer likely values of the weights given some labeled data.
\begin{equation}
  \label{rvm.bayes}
P(\beta | {\bf x}, t) \propto P(\beta) P(t | {\bf x}, \beta)
\end{equation}
In this expression, we have introduced an extra probability distribution, $P(\beta)$,
which is our prior belief in the values of the weights.  If we merely wished to
perform classical GLM training, we would have no preference for any particular
value of the weights, and would encode this by providing a very broad (``non-informative'')
prior.  However, as discussed previously, we believe that simple models are more
likely to make useful generalizations.  A preference for simplicity can be
encoded using an Automatic Relevance Determination prior \cite{mackay.ard}.  In this
case, we introduce an additional vector of parameters,
$\alpha$.  Each element of the $\alpha$ vector controls the width of the prior
over the corresponding weight:
\begin{equation}
  \label{rvm.prior}
P(\beta) = \sum_{m=1}^M {\mathcal G}(\beta_m | 0, \alpha_m^{-1})
\end{equation}
To include these new $\alpha$ parameter in the inference process, we also need
to specify a hyperprior over values of $\alpha$.  For the RVM, a very broad gamma
distribution is used -- a standard choice for non-informative priors over
Gaussian width parameters -- but the precise choice is of little importance so
long as it is sufficiently broad.

Considering just a single basis function, there are two possibilities:
\begin{itemize}
\item{The basis function provides additional information about the specified
  classification problem.  When its weight is set to some non-zero value, the
  amount of misclassified training data is reduced.  This increases the value of
  equation \ref{rvm.likelihood}, and therefore the probability of that model given the data.}
\item{If the basis function provides no information because it is
  irrelevant to the problem,
  there is no value of the weight that will lead to a significant increase in the
  likelihood.  At this point, the prior term in the model comes into play: by setting
  the $\alpha_m$ parameter to a large value, the prior distribution $P(\beta_m)$ becomes sharply peaked
  around zero.  By then setting $\beta_m$ to zero, the posterior probability
  of the model is maximized.  Similarly, when two basis functions offer redundant
  information, the posterior is maximized by using only one of them in the model.}
\end{itemize}
When a basis function has a sufficiently high $\alpha$, it can be marked as
irrelevant, and removed from the model.  As a result, the RVM will learn simple
models even when presented with a large starting set of basis functions.

In this discussion, we have neglected the practical issues of evaluating equation
\ref{rvm.bayes} to find good values for the model parameters.  A number of strategies can
be used, including Monte-Carlo simulation, maximum likelihood estimation \cite{tipping.rvm}, and
variational inference \cite{mackay.ensemble}.  While we have experimented with all these methods,
the training library which we use used when developing all the models described
here is based on the variational approximation given in \cite{bishop.vrvm}.  This
training method involves iterative update of the model parameters, and it is
possible to remove irrelevant basis functions during the course of the training,
so the training process runs progressively faster as it converges towards the
optimal solution (and corresponding minimal set of basis functions).

\subsection {Sampling-RVM}

In principle, we would like to provide as little prior knowledge as possible
about which feature spaces to consider when solving a problem, instead allowing
a sparse trainer such as the RVM to pick freely from a wide range of possible
basis functions.  Unfortunately, computational costs rapidly become
significant.  The progressive simplification of the problem means that it is
difficult to quantify the full computational complexity since the size of the
matrices varies from cycle to cycle.  However, scaling is quite substantially
worse than linear.

Here, we suggest a pragmatic approach to handling large sets of candidate basis function.
A working set is first initialized with a subset of basis
functions, picked at random from the complete pool of candidates.  The trainer is run as
previously described, and as before some $\alpha$ values increase to a level
such that it is possible to remove the associated basis functions from further
consideration. Once the size of the working set drops below a designated
`low water mark', additional basis functions are added from the pool. At this
point, all $\alpha$ and $\beta$ values are reinitialized and training
continues. In this way, the working set fluctuates between high and low water
marks until the pool is exhausted, at which point the trainer continues to run
until the weights and priors no longer change significantly between cycles
({\it i.e.} convergence) to give a complete model.

\begin{figure}[!bth]
\begin{center}
\includegraphics[scale=0.4]{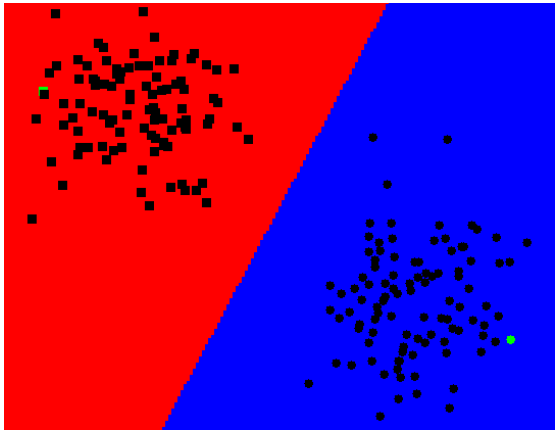}
\caption{Example of a data set used for testing of training speed.  The two points
  chosen as centers for radial basis functions are highlighted in green.}
\label{speed.test}
\end{center}
\end{figure}

To evaluate the performance benefits of this method, we considered a simple 
classification problem in Cartesian 2-space.  This involved two equal-sized
classes of points sampled from two Gaussians, with a substantial space between
them.  Basis functions were generated in an ``SVM-like'' manner, with a radial basis
function centred on each point in the training data.  Thus, the size of the basis-value matrix
of the RVM trainer increases with the square of dataset size.
As shown in figure \ref{speed.test}, an optimal model requires
only two basis functions, and gives 100\% classification accuracy on the training
data.

\begin{figure}[!bth]
\begin{center}
\includegraphics[scale=1.0]{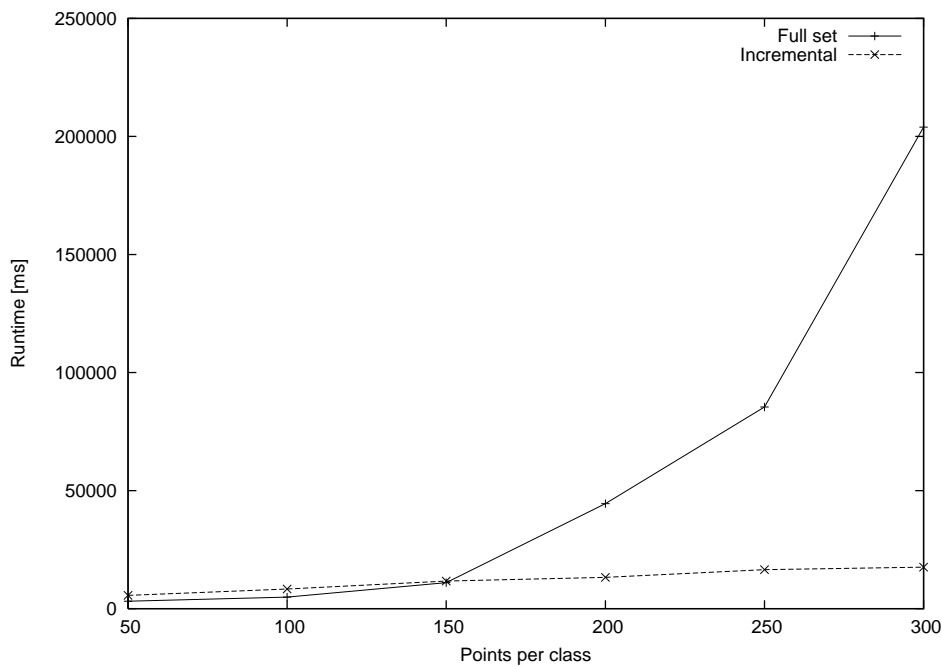}
\caption{Training time vs. problem size for full-set and incremental RVM learning.}
\label{incremental.speed}
\end{center}
\end{figure}

A range of problem sizes were tested, from 50 points in each class (i.e. 100 basis
functions) to 300 points.  We compared the full-set training method with
the incremental method, using a high water mark of 20 basis functions and a low
water mark of 15.  Timings are shown in figure \ref{incremental.speed}.  
For small problems, the incremental training approach is in fact significantly
slower, due to the increased number of cycles and the need to periodically restart
the training process as more basis functions are added.  However, the scalability
of the method to large problems is substantially better.

It is interesting to note that, while in all cases all the training data was
correctly classified, the full-set training runs with 300 points did not learn
the simple 2-basis function model seen with smaller training sets, but
instead appeared to converge before all unnecessary basis functions had been removed.
However, the incremental trainer consistently learned a fully sparse model for the same
dataset with no complications.  We believe that this indicates the limits of
either the linear algebra routines used by the trainer, or the
precision of the standard `double' floating point mathematics available on most
computers.  Errors caused by lack of
precision (often described as numerical instability) are a common problem in
numerical computation, and large linear algebra computations are a frequent source
of trouble.  By keeping the size of individual computations down, using the
small-working-set variant of the RVM neatly sidesteps these problems and makes
it possible to tackle very large problems efficiently.


Having made the step toward the approach of gradually sifting through a large
set of basis functions, it is only a small additional step to consider an
implementation which generates new basis functions on the fly during the training
process.  In principle, this can allow exploration of infinite sets of basis
functions.  This is particularly feasible in cases where there are clear correlations
between basis functions.  For instance, a radial basis function with centre
(10.0, 10.2) and variance 10 will have outputs that are highly correlated to
a second radial basis function with centre (10.0, 10.0) and equal variance.  This
means that if the first function is found to be informative for modeling a 
particular problem, it is likely that the second will also be effective.  Therefore,
after some period of training, it is likely to be worthwhile proposing more
candidate basis functions which are correlated to those currently in the
working set.

In many practical applications, including the sequence models described here,
a significant proportion of total run time may be spent evaluating the basis
functions.  However, since the evaluation of each basis function for each item
in the training set is an independent operation, this part of the process is
easily parallelized on both individual multiprocessor computers and clusters.

\subsection{Position-weight matrices}

When studying biological sequences, it is often natural to look for significant
short motifs.  In many cases -- for example, gene promoters and exonic splice
enhancers -- these motifs act as preferred binding sites for specific proteins.
It is rare, however, that there is only one sequence which will bind a particular
protein: normally, there are at least some positions in the motif where more
than one base is acceptable.  Given a large number of examples, a good way to represent knowledge
about a motif is a Position Weight Matrix (PWM) \cite{bucher.wms}. This is
a matrix where each element represents the probability of a given nucleotide
occurring at a particular position in the sequence.  In other words, each column
of the matrix is a probability distribution over the DNA alphabet.  A PWM can be
viewed as a probabilistic model of a fixed-length sequence:
\begin{equation}
W(s) = \prod_{i=1}^{|W|} W_i(s_i)
\end{equation}
Where $|W|$ is the length of the PWM, and $W_i$ is the probability distribution
represented by the
$i$'th column of the matrix.  The higher the probability of a given sequence
under this model, the more similar the sequence is to the ``ideal'' motif, and
therefore the higher the probability that it will function as a binding site.  
This is the zeroth order model: each position in the sequence is assumed
to be independent of all the others.


It should be pointed out that the assumption of bases being independent in a motif
is unlikely to be true in most cases.  When proteins bind to DNA, they often
significantly deform it, so mechanical properties of the DNA other than the actual
base sequence will be significant -- in particular, the flexibility of the double
helix.  This property is determined largely by interactions between neighbouring
base pairs.  Models which take these, and perhaps also longer range interactions, into
account can be expected to better predict the binding of a protein to a given
sequence \cite{barash.depend}.  However, since non-independent models have
many more parameters than simple PWMs, they require more example sequences to
learn effectively.

\section{Eponine RVM models}

\subsection{Eponine Anchored Sequence}

The Eponine Anchored Sequence (EAS) model is a system for analyzing the sequence
around some particular interesting point (the anchor point).  A number of important
sequence analysis questions, such as prediction of transcription start sites
and splice sites, are best expressed in terms of classifying individual points in
the sequence.  In practice, the data is presented to this model as a large piece
of sequence data and an integer
defining the anchor point under consideration.  When searching for features
in bulk sequence, the same sequence is presented many times while scanning the
anchor point along its length.  The basic element of an EAS model is the
positioned constraint (hereafter, PC).  This consists of:
\begin{itemize}
\item{A preferred sequence motif, defined as a position-weight matrix.}
\item{A probability distribution over integer offsets relative to the anchor
point, which defines the expected localization of the motif.  In the work presented here, these
distributions are always discretized Gaussians ({\it i.e.} the result of integrating
the Gaussian probability density function over unit intervals).  Gaussians were
chosen because of their familiarity, and a smooth shape that made Gaussian-based
models less prone to overfitting than functions with abrupt changes, such as square
waves.  However, any distribution over integers could in principle be used here.}
\end{itemize}
To obtain a score for a PC on a given piece of data, the program scans over all positions in the sequence which are
assigned a non-infinitesimal probability by the chosen position distribution. 
For each position, the probability of the sequence motif starting at that
position being emitted by the chosen weight matrix is evaluated.  The final score
is given by:
\begin{equation}
\phi(C) = \frac{1}{|W|} \log \sum_{i = -\infty}^\infty P(i) W(C, i)
\end{equation}
where $C$ is a genomic context, $P$ is a position distribution, and
$W(C, i)$ is a DNA weight matrix probability for offset $i$ relative to the
anchor point of  $C$.  Note the division by $|W|$, the number of columns in the
weight matrix ({\it i.e.} the length of the sequence motif which it defines).  This is
important, since this method allow motifs with a wide range of lengths -- with
the trainer implementation described here, the length varies between 2 and 20
columns.  The RVM trainer has a weak bias towards selecting basis function with
higher absolute magnitudes.  Normalizing the scores allows unbiased selection between
motifs of different lengths.  It is somewhat analogous to the whitening process
often used to pre-process data for SVM classifiers, where all the training vectors
are normalized to constant length \cite{svm.book}.

\begin{figure}[!bth]
\begin{center}
\includegraphics[scale=0.66]{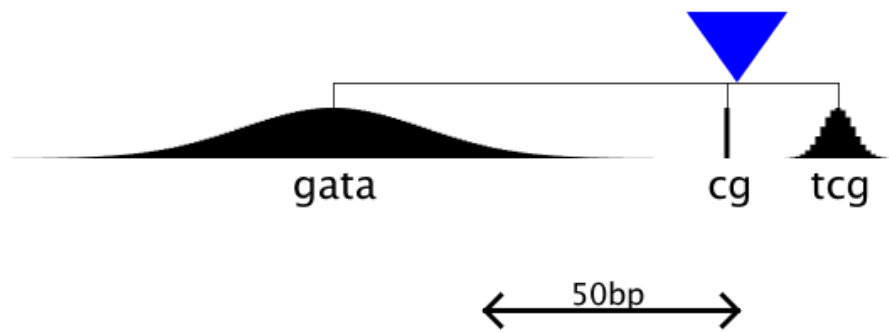}
\caption{Example schematic architecture of an Eponine Anchored Sequence model.}
\label{eas.example}
\end{center}
\end{figure}

A single PC describes an individual sequence motif and its relationship to a
point in a sequence, but a set of them can be combined to describe more complex
structures.  Figure \ref{eas.example} shows a schematic of a model
combining three positioned constraints.  If the final output score is defined as a weighted sum of individual PC
scores, the combined model is a generalized linear model over genomic contexts,
with the PCs as basis functions.  Therefore, it is possible use the Relevance
Vector Machine method to reduce a large set of candidate PCs down to a sparse
model containing a small, informative subset.

While this description of the model architecture emphasizes detection of
single, well-defined ``words'' in the sequence, since the overall PC score is based
on a sum of weight matrix scores across a region, it is also possible to represent
general compositional biases of a region by picking a PC with a short weight matrix
and a very broad position distribution.  It is also quite acceptable for the 
distributions of two or more PCs to overlap.  Since
arbitrary weights are assigned in the training process to fit the model to the training
data, any issues with double-counting of a particular piece of information are corrected
automatically

The space of potentially interesting PCs is extremely large.  Even taking a
highly simplified view, restricting the constraints to simple motifs rather than
weight matrices and the position distributions to Gaussians of a constant width,
there are over one million PCs representing six-base motifs with distributions
centred at positions in the range [-250:50] relative to the anchor point. This
space it too large to search exhaustively.  Of course, in practice the EAS
framework allows for an infinite (limited only by numerical precision on the
probability values) number of PCs.  Fortunately, as in the case of radial basis
functions, the space of possible basis functions for this model is highly correlated. 
Making a small change to, say, one of the probabilities in a weight matrix will
give a second PC whose output on a given sequence is correlated with the first. 
For this reason, exploring regions of a conceptual ``PC-space'' in the
neighborhood of constraints which have already proved to be informative is
likely to reveal even more informative constraints.

Taking advantage of these correlations, EAS models can be trained using an RVM
with a sampling strategy to create new basis functions.  When the size of the
working set falls below the low water mark, the trainer selects a sampling
strategy at random from the following set:
\begin{itemize}
\item{Constructing a new PC, not based on the current set.  This is performed by
the following algorithm:
\begin{enumerate}
\item{First, select a context at random from the training set (either
positive or negative, without bias).}
\item{Pick some point relative to that context's anchor point.}
\item{From that point, take a sequence motif of between 3 and 6 bases in length,
and construct a weight matrix which optimally matches that consensus sequence,
but includes some degree of uncertainty.}
\item{Construct a PC using the newly selected weight matrix with a Gaussian position
distribution of random width, centred at the position at which the motif
was originally found.}
\end{enumerate}
Obviously, PCs selected in this way will strongly match the training example from which
they were originally derived.  This is closely analogous to selecting radial basis functions centred
on points in the training data set.
}
\item{Selecting an existing PC and adjusting the emission spectrum of one column
of its weight matrix, by sampling from a tightly-focused Dirichlet distribution
centered on the current values.}
\item{Adding an extra column to either the start or the end of an existing weight
matrix, up to a maximum number of columns (in this case 20).}
\item{Removing the start or end column from an existing weight matrix, down
to a minimum of two columns}
\item{Adjusting the width parameter of a Gaussian position distribution}
\item{Adjust the centre position for a Gaussian position distribution}
\end{itemize}
For an initial period of 200 cycles, only the first sampling rule (creation of
novel basis functions) was used.  After this point, the full range of sampling strategies
were available, and training proceeded by a mixture of sampling and introduction of
novel PCs.

This model was originally developed for the purposes of transcription start
site prediction, and has been shown to perform well for that purpose, at
least in mammalian genomes \cite{down.epotss, down.thesis}.  Subsequently, it
has been used for predicting transcription termination sites [A. Ramadass, personal
communication].

\subsection{Eponine Windowed Sequence}

The Eponine Windowed Sequence (EWS) model is designed by analogy to the EAS model,
but rather than targeting individual points in the sequence, it is designed
to classify small regions or windows of a sequence, based purely on their own
sequence content.  A naturally windowed classification problem would be to distinguish
exons from intergenic sequence.  Once again, the sensors are position-weight matrices, but each
weight matrix is scanned across the entire region.

Normalizing for the length of the sequence being inspected and
the size of the PWM, the basis functions of the model take the form:
\begin{equation}
  \label{ews.basis}
\phi(S) = \frac {4^{|W|}} {|S|-|W|+1} \sum^{|S|-|W|+1}_{i=1}W(S^{i+|W|-1}_i) 
\end{equation}
where $W(s)$ is the probability that sequence $s$ was emitted by weight matrix
$W$, and $S^j_i$ denotes a subsequence from $i$ to $j$. 

An initial set of basis functions is proposed by taking all possible DNA motifs
of a specified length (typically 5) and generate weight matrices which recognize
these motifs.  When new basis functions are needed, one of the following sampling moves
is used to generate a variant of a basis function currently in the working set.
\begin{itemize}
\item{Generate a new weight matrix in which each column is a sample from a
  Dirichlet distribution with its mode equal to the weights in the corresponding
  column of the parent weight matrix.}
\item{Generate a new weight matrix one column shorter than the parent by removing
  either the first of the last column.}
\item{Generate a new weight matrix with an extra column at either the start or
  the end, biased in favour of a random base.}
\end{itemize}

We have used the EWS model to explore common motifs in regions of non-coding
sequence conserved between the mouse and human genomes, as discussed in
\cite{down.learncompare}.  We have also successfully trained this model on
windows of sequence upstream of annotated genes, in which case it works successfully
as a promoter predictor, although the EAS model is generally superior for this
purpose since it is able to make predictions of the exact locations of transcription
start sites, rather than merely highlighting broad regions likely to have promoter
activity.

\subsection{Convolved Eponine Windowed Sequence (C-EWS)}

The C-EWS model is an extension of EWS to capture larger-scale patterns in sequences.  In this case,
each basis function defines a `scaffold' consisting of one or more PWMs, each
with an associated position distribution relative to a scaffold anchor.  In principle,
distributions such as the discretized Gaussian extend to infinity, but in practice
it is reasonably to apply some cut-off: for instance, only considering the portion
of the distribution which includes 99\% of the total probability mass.  The probabilities
of all points outside this region are assumed to be infinitesimal and ignored.  Now
that the distributions have finite size, for a given scaffold there is a pair
of integers, $n$ and $m$, such that when the scaffold anchor is placed
in the interval [$n$ : $m$], the non-infinitesimal parts of all the
position distributions fall entirely within the length of a particular target sequence.
A score for scanning this scaffold across the sequence can be given by
\begin{equation}
  \label{cews.basis}
\phi(S) = Z\sum^{m}_{i=n}\biggl(\prod_{k=1}^K\bigl(\sum_{j=-\infty}^{\infty}P_k(j)W_k(S_{i+j}^{i+j+|W_k|})\bigr)\biggr)
\end{equation}
where $P_k$ is the $k$'th position distribution and $W_k$ is the
$k$'th weight matrix in the scaffold.  $Z$ is the normalizing constant:
\begin{equation}
Z = \frac{4^{\sum_{k=1}^K |W_k|}}{m - n + 1}
\end{equation}
For the case when the scaffold only contains one PWM with a narrow distribution,
the results will be the same as those from equation \ref{ews.basis}.  So this can be
considered a direct extension to the original EWS model that can capture information
about sets of motifs with correlated positions.  Since the scaffold scores are only
evaluated in the regions where the whole scaffold fits onto the sequence window,
there is a risk of introducing edge effects.  A possible future solution
to this would be to use windows that are a little larger than the actual region,
of interest, and use ``soft boundaries'' where scores from the edges of the window
are given less weight than those at the centre.  To train scaffold-based models,
all the sampling rules for the EWS model are used, together with some additional
rules for constructing and adjusting scaffolds:
\begin{itemize}
\item{Combine the sets of motifs from two scaffolds, with randomly chosen offsets
between the two (up to some maximum number of weight matrices per scaffold).}
\item{Take a scaffold with two or more PWMs and return the scaffold with one
of those PWMs (picked at random) removed}
\item{Alter the position or width of one of the relative position distributions in
a scaffold.}
\end{itemize}
To date, we have only investigated scaffolds with small numbers of PWMs (up to
a maximum of 3).  If more complex scaffolds were used, it might be appropriate
to consider additional sampling rules which combined a subset of PWMs from two
parent scaffolds: this would be analogous to the use of `cross-over' rules in
the field of genetic algorithms.

The C-EWS model can be applied in any situation where EWS can be used.  It should
never give significantly worse results, but is better able to capture longer-range
features in a sequence: for instance binding sites for two separate proteins which
interact with one another (and therefore have binding sites close to one another).
This model scheme has recently been applied successfully to the discovery of
exonic splice enhancer motifs in coding exon sequence [B. Leong, manuscript in
preparation].

\section{Conclusions}

The Sampling-RVM learning systems presented here are powerful strategies for
analyzing sequence data.  All these methods have been successfully applied
in the past and we hope they will be applied more widely and on new systems
in the future.  We continue to develop and experiment with these methods,
with a particular interest in promoter and transcription start site research.

It should be noted that in all cases, it is in principle possible to use any
of these models to distinguish between more than two classes of object -- for example,
a set of promoter training data could be split into several classes based on the
tissues in which they are active.  However,
the variational RVM trainer which we have used here relies on an approximation
which is only valid when considering two-class problems.  We have developed
a prototype RVM implementation which uses Metropolis-Hastings sampling to
perform the Bayesian inference, and this has successfully completed three-way
classification problems, albeit at the cost of increased computational requirements.

A key part of the strength of these approaches is the use of the Relevance
Vector Machine as an extremely flexible method for training sparse models.
The RVM has already been applied in the classification of microarray gene expression
data {pocock.thesis}, and we encourage its use in other areas of bioinformatics
and beyond.

\section{Availability}

The RVM training library and Eponine sequence models are implemented in the
Java programming language, using the COLT and BioJava libraries.  Java source code is
available to all interested researchers on request from TD.

\section{Acknowledgments}

Thanks to Matthew Pocock for many interesting discussions on machine learning
in bioinformatics.  TD would like to thank the Wellcome trust for funding.

\end{document}